\documentclass{article}
\usepackage{amssymb,amsmath}

\usepackage{graphicx}   
\usepackage{verbatim}   
\usepackage{color}      
\usepackage{subfigure}  
\begin{document}
\topmargin=-0.4in \oddsidemargin=-0.2in

\textheight=8.8in \textwidth=6.8in

\begin{center}\textbf{\Large  Toric manifolds for Flux compactification}
\vspace{0.3cm}\end{center}

\begin{center}{\large Sujan P. Dabholkar$^{\star}$ }\end{center}{\large \par}

\begin{center} SCGP/YITP, Stony Brook University, \\
 Stony Brook, NY 11794 USA\\
$^{\star}$sujan.dabholkar@stonybrook.edu
\end{center}

\begin{center}\vspace{0.3cm}\end{center}

\begin{abstract}
We study how to use smooth, compact toric varieties for supersymmetric $AdS_{4}$ flux compactifications using tools of SU(3) structures, similar to $\mathbb{CP}^3$ solution. A non-vanishing globally well defined complex 3-form plays a key role in such constructions. Necessary topological conditions associated with it will be understood to put constraints on large class of these manifolds for supersymmetric flux compactification. Local analysis of SU(3)-structure is carried out, which might help to explore more flux vacua.
\end{abstract}




\eject
\tableofcontents


\section{Introduction}

\label{introduction}

If string theory describes real-world physics, we must compactify it from ten dimensions
to four dimensions such that we have four uncompactified external space-time dimensions and six compact internal dimensions. If we are working with supergravity which is the low-energy limit of string theory, one has to be careful such that the compactifications constructed are in a regime where the supergravity description is valid (large volume limit). 

The widely studied supersymmetric compactifications are using Calabi-Yau manifolds as the internal space. As soon as one turns on background fluxes to obtain supersymmetric vacua, the internal manifold cannot be Calabi-Yau\footnote[1]{In GKP, one can turn on fluxes and internal manifold is conformal Calabi-Yau.}. The main idea of string compactification with fluxes is discussed in \cite{Grana:2006ab, Douglas:2006ab, Lust:2006a}. By now several supersymmetric AdS$_4$ flux vacua are known. Constructing a string vacuum with a positive cosmological constant poses lot of difficulties, such as use of orientifolds to evade No-go theorems. For Type IIB solutions, one needs to play with non-perturbative effects for Kahler moduli potential. In case of type IIA compactifications, one can turn on fluxes and all geometric moduli fields can be stabilized classically with O6-planes where supergravity description is valid\cite{DeWolfe:2005ab}. Recently, some issues with such moduli stabilization with O6-planes are discussed in \cite{Sethi:2012ab}. Lot of progress is happening in Type II/Heterotic flux compactification and uplifting to dS vacua, but here we will focus on string vacua with negative cosmological constant with some supersymmetry.

Toric varieties have played an important role in string compactifications and mirror symmetry: as Calabi-Yau manifolds are embedded in them as hypersurfaces. Once fluxes are turned on, the three-dimensional smooth, compact toric varieties can be used for compactifications with the help of SU(3)-structures, instead of considering them as embedding spaces.  In string theory/M-theory compactifications, $\mathbb{CP}^3$ has played a great role in constructing explicit examples. We will study how to use more general smooth toric manifolds for flux compactification following the procedure given in \cite{Gaiotto:2009a,Larfors:2010lt}.

 We will mainly deal with the symplectic quotient description of smooth toric variety and the construction of SU(3)-stucture on it.  Massive type IIA vacua with $\mathbb{CP}^3$ was obtained by considering  $\mathbb{CP}^3$ as a twistor fibration of $S^2$ on $S^4$ with unusual almost complex structure which is not integrable. In section 4, we will discuss about topological restrictions for carrying out the procedure of changing almost complex structure similar to $\mathbb{CP}^3$ on smooth toric manifolds. This puts the constraint on large number of toric manifolds in order to use them for flux compactifications. The first Chern class is commonly used to study string compactification, in section 4.3, we study Top Chern class and propose its
use in the compactifications with SU(3)x SU(3) or strict SU(3) structure manifolds.

In section 5, we will carry out the local analysis of SU(3) structure conditions and will show that we have many parameters to change the torsion classes associated with SU(3)-structure, this leads to the possibilty of obtaining more Type IIA flux vacua using smooth, compact toric manifolds. Also, one can use the procedure for compactification of Heterotic theories \cite{Gray:2012,Frey:2005}.

\section{Supersymmetry and G-structures}\label{rigid}

Supersymmetry requires the existence of nonvanishing, globally well defined spinor on the internal manifold.  This condition puts some topological restrictions on the internal manifold. This is very well understood and various cases are known for supersymmetric vacua \cite{Grana:2005a}. Numerous cases for Type II supergravity with such restrictions are known by now, for our purpose some useful cases are mentioned in \cite{Tomasiello:2007a,Bovy:2005,LT:2009a,LT:2009b,LT:2004} . We mainly focus on strict SU(3) and dynamic SU(3)$\times$SU(3) structures. SU(3) structure manifolds play key role in $\mathcal{N}=1$ compactification of Heterotic strings and Type IIA compactifications.

\subsection{Strict SU(3) structure}\label{rigid}

In this subsection, we discuss the idea of SU(3) structure\cite{Salamon:2002a}.\\
Manifolds with SU(3) structures admit one globally defined, nonvanishing spinor $\eta$. This structure can be understood through (J,$\Omega$) forms. J is a real (1,1) form and $\Omega$ is a complex (3,0) form such that $J \wedge \Omega=0$ and $i\Omega \wedge \bar{\Omega}=\frac{4}{3}J^{3}\neq 0$.  In terms of spinors, one can write $J_{ab}=i\eta^{\dagger}_{-} \gamma_{ab}\eta_{-}$ and $\Omega_{abc}=\eta^{\dagger}_{-} \gamma_{abc}\eta_{+}$.
\begin{description}
 \item[1.] $dJ=\frac{3}{2}Im(\bar{W_1}\Omega)+W_{4}\wedge J+W_3$ 
 \item[2.]$d\Omega=W_{1}J^2+W_{2}\wedge J+\bar{W_5}\wedge\Omega$
\end{description}

Here $W_{1}$ is a complex scalar, $W_{2}$ is a complex primitive (1,1) form, $W_{3}$ is a real primitive (1,2)+(2,1) form, $W_{4}$ is a real one form and $W_{5}$ is a complex (1,0) form.

If $W_{1}=W_{2}=0$, then the manifold is complex and $W_{1}=W_{3}=W_{4}=0$ will be a symplectic manifold. All $W_{i}=0$ lead to Calabi-Yau.


\section{G-structures on smooth, compact Toric manifolds}\label{rigid}

The toric manifold $M_6$ is K$\ddot{a}$hler and admits a global U(3) structure naturally.

\subsection{SU(3)-structure}\label{rigid}

In this section, we discuss a general procedure for constructing string compactifications
on smooth toric varieties via a method for producing SU(3)-structures\cite{Gaiotto:2009a},\cite{Larfors:2010lt}.

Consider the quotient description of the toric variety(Real dimension $2d$)\cite{Denef:2008a}. If $\{z_{i}, i=1,..,n\}$ are the holomorphic coordinates of ambient space $\mathbb{C}^{n}$ such that toric action is $\{z_{i}\rightarrow e^{iQ^{a}_{i}\alpha_{a}z^{i}}\}$. Then the toric variety is described as 
\begin{equation}
\mathcal{M}_{2d}=\{z^{i}\in \mathbb{C}^{n}|\sum_{i}Q^{a}_{i}|z^{i}|^2=\xi^{a}\}/U(1)^{s}
\end{equation}

The toric variety has the induced real form $\tilde{J}_{FS}$ and a complex d-form $\tilde{\Omega}_{FS}$.
\begin{equation}
\tilde{\Omega}_{FS}=(det(g_{ab}))^{-1/2}\Pi_{a}\imath_{V^{a}}\Omega_{\mathbb{C}}
\end{equation}
Here $\Omega_{\mathbb{C}}$ is a holomorphic top form on the ambient space. It is easy to see that $\tilde{\Omega}_{FS}$ is vertical and regular with no poles. 
Lets start with a (1,0) form K with respect to complex structure on the ambient space $\mathbb{C}^n$ and the holomorphic vector fields generating the $U(1)^{s}$ action, $V^{a}=\sum_{i}Q^{a}_{i}z^{i}\partial_{z_{i}}$ such that K satisfies following conditions.

    \begin{description}
      \item[(a)] K is vertical. $\imath_{V^{a}}K=0$.
      \item[(b)] It has a definite $Q^{a}$-charge.
		$\mathcal{L}_{ImV^a}K=q^{a}K$ where $q^{a}=\frac{1}{4}\sum_{i} Q^{a}_{i}$. This condition is required to have well-defined 3-form on $\mathcal{M}_{6}$.
      \item[(c)] K is nowhere-vanishing. This condition needs some special attention.
    \end{description}
Conditions (a) and (b) tell us that K is not well-defined on $\mathcal{M}_{2d}$. But the local SU(2)-structure comprising of a real two-form J and a complex two-form $\omega$ can be obtained using K.
\begin{eqnarray}
\omega &=-\frac{i}{2}K^{*}\cdot \tilde{\Omega}_{FS}\\
j&=\tilde{J}_{FS}-\frac{i}{2}K\wedge K^{*}
\end{eqnarray}
We note that the construction for SU(3) structure suggested in \cite{Larfors:2010lt} using local SU(2) structure is obtained using j, $\omega$ and K. We argue that condition 3: K is nonvanishing everywhere is not required, only topological condition we need is $c_1(\mathcal{M_{6}})=0$. (j,$\omega$, K) can have zeros or poles. 
\begin{eqnarray}
J&=\alpha j-\frac{i\beta^{2}}{2}K\wedge K^{*}\\
\Omega &= e^{i\gamma}\alpha \beta K^{*}\wedge \omega
\end{eqnarray}
Here $\alpha$,$\beta$ and $\gamma$ are real, gauge invariant, nowhere-vanishing functions and $\tilde{\Omega}_{FS}$ is a complex d-form on $\mathcal{M}_{6}$. $\Omega$ obtained with such K is well-defined on 6-manifold. 


\subsection{comment on static SU(2)-structure}\label{rigid}

Let's look at what happens in SU(2)-structures. The manifold with static SU(2)-structure admits two nonvanishing globally-defined spinors $\eta_{i}, i = 1, 2$, that are linearly independent, orthogonal at each point. From supergravity point of view, such manifolds in general lead to $N=4$ SUGRA in 4 dimensions\cite{Bovy:2005}\cite{Louis:2009}. For Type IIA point of view, there is no solution on manifolds with static SU(2)-structure.

The SU(2)-structure on 6 manifolds is characterized by a non-vanishing complex one-form K, a real two-form J and a complex two-form $\Omega$. For our purpose, consider the one-form K, it follows
\begin{align}
&K\cdot K=0\nonumber\\ 
&K^{*}\cdot K=2\nonumber\\
&K_{j}=\eta^{c}_{2}\gamma_{j}\eta_{1}
\end{align} 
In order to study SU(2)-structures on smooth, compact toric manifolds, we need to have a nonvanishing section of cotangent bundle. It is known in mathematics\cite{BoTu:1982} that if section of a tangent bundle (E) is nonvanishing, the Euler class, $e(E)=0$. We know that top Chern class of a smooth, complete toric variety is $c_{n}=|\Sigma(n)|[pt]$, where $[pt] \in H^{2n}(M, \mathbb{Z})$. Hence, we cannot have static SU(2)-structure on smooth, compact toric varieties. Thus we cannot obtain $\mathcal{N}=4$ supergravity compactification on such manifolds.


\section{Topological conditions for Toric compactifications}\label{kk}

\subsection{$\mathbb{CP}^3$ case}\label{rigid}

Consider $\mathbb{CP}^3$ as a twistor fibration on $S^{4}$ with $S^{2}$ as a fiber. $$S^{2} \hookrightarrow CP^{3} \rightarrow S^{4}$$ Naturally one can consider almost complex structure. $S^{2}$ is diffeomorphic to $\mathbb{CP}^1$.
\[ \mathcal{I} = \left( \begin{array}{ccc}
\ I_{2} & 0 \\
0 & I_{4}  \end{array} \right).\] 
This almost complex structure is integrable. Locally $TM_{\mathbb{C}}=TM_{\mathbb{R}}\otimes\mathbb{C}=T^{(1,0)}M\otimes T^{(1,0)}M$.  We know that locally $T^{(1,0)}{\mathbb{CP}^3}=T^{(1,0)}{\mathbb{CP}^1}\oplus \xi$.\\
Thus, for integrable almost complex structure, using $c(T^{(1,0)}{\mathbb{CP}^3})=c(T^{(1,0)}{\mathbb{CP}^1})c(\xi)$, if g is the element in $H^{2}(\mathbb{CP}^1)$, we have \\
\begin{equation}
(1+4g+6g^2+4g^3)=(1+2g)(1+c_{1}(\xi)+c_{2}(\xi))
\end{equation}
We get $c_{1}(\xi)=2g$ and $c_{2}(\xi)=2g^2$.\\

To obtain non-integrable almost complex structure, we will consider
\[ \mathcal{I} = \left( \begin{array}{ccc}
\ -I_{2} & 0 \\
0 & I_{4}  \end{array} \right).\] 
 
Let's study what happens when we make the change in $I_{2}$. we have
\begin{equation}
(1-2g)(1+2g+2g^{2})=(1+c_{1}^{new}+c_{2}^{new}+c_{3}^{new}) 
\end{equation}
We get $c_{1}^{new}=0$ and $c_{2}^{new}=-2g^2$ and $c_{3}^{new}=-4g^3$.\\ 

Even though we have the same real tangent bundle, due to the choice of almost complex structure, we have modified the complex tangent bundle. It was known to mathematicians that this change leads to vanishing 1st chern class, but it is important to notice that $c_{top}$ does not vanish, which is anyway expected as it is equal to euler class of M which doesn't depend on the choice of almost complex structure. It just picks up a sign based on orientation. This new almost complex structure leads to $c_{1}=0$ and there is a globally defined 3-form.


\subsection{Smooth, compact Toric varieties}\label{rigid}

In this section, we see how constrained such change in almost complex structure is for smooth, compact Toric varieties.

Consider a smooth, compact Toric variety $\mathcal{M}_{6}$,  with a four-two split of tangent bundle, not necessarily restricted to twistor space or product manifold. This is obtained using the almost product structures\cite{Grana:2005a}. The tangent bundle at a point can be split into two parts.
Following previous section, $T^{(1,0)}{\mathcal{M}_{6}}=T^{(1,0)}{\mathcal{M}_{2}}\oplus \xi$.\\
Thus, for such almost complex structure, using $c(T^{(1,0)}{\mathcal{M}_{6}})=c(T^{(1,0)}{\mathcal{M}_{2}})c(\xi)$, we have \\
\begin{equation}
c_{1}(T^{(1,0)}{\mathcal{M}_{6}})=c_{1}(T^{(1,0)}{\mathcal{M}_{2}})+c_{1}(\xi^{(1,0)})
\end{equation}

Now let's understand the flip in the almost complex structure which leads to $c^{new}_{1}(T^{(1,0)}{\mathcal{M}_{6}})=0$.
\begin{equation}
0=-c_{1}(T^{(1,0)}{\mathcal{M}_{2}})+c_{1}(\xi^{(1,0)}). 
\end{equation}
From (10) and (11), we get $c_{1}(T^{(1,0)}{\mathcal{M}_{6}})=2\times c_{1}(T^{(1,0)}{\mathcal{M}_{2}})$. In terms of divisors, $c_{1}(T^{(1,0)}{\mathcal{M}_{6}})=\sum_{i}D_{i}$.\\ 

Thus, in order to get vanishing 1st chern class, it is important to notice that the four-two split satisfies above condition. Then one can do the compactification of String theory on smooth, compact Toric variety.

We should see this condition with an example: $\mathbb{CP}^1$ bundle over $\mathbb{CP}^2$, discussed in Denef's review\cite{Denef:2008a}. It can be described as 

\[
 \mathcal{M}_{6}=\left\{ x \in \mathbb{C}^{5}|
  \begin{array}{l l}
      |z_{1}|^{2}+|z_{2}|^{2}+|z_{3}|^{2}-n|z_{4}|^{2}=\xi_{1} & \quad \\
      |z_{4}|^{2}+|z_{5}|^{2}=\xi_{2} & \quad \\
  \end{array} \right\}/U(1)^{2}
\]

In this case, n accounts for the "twisting" and action of $U(1)^2$ is given by
\begin{center}
$(z_{1},z_{2},z_{3},z_{4},z_{5}) \rightarrow (e^{i\phi_{1}}z_{1},e^{i\phi_{1}}z_{2},e^{i\phi_{1}}z_{3},e^{i(\phi_{2}-n\phi_{1})}z_{4},e^{i\phi_{2}}z_{5})$
\end{center}

The divisors for this smooth toric manifold are $D_{1}=D_{2}=D_{3}$ and $D_{4}=D_{5}-nD_{1}$. The first chern class is given by $c_{1}=(3-n)D_{1}+2D_{5}$. For $n=3$, $c_{1}(M)=2D_{5}$, thus in this case, we can obtain the vanishing first cherm class, by choosing proper divisor and corresponding sign flip. For $n=2$, $c_{1}(M)=D_{1}+2D_{5}$, in this case, we cannot obtain the vanishing first cherm class. In general, all odd twistings are allowed.
For a complicated case, it is given by $c_{1}(M)=\sum_{i}D_{i}$. Note that in this discussion, $D_{i}$ represents Poincare dual associated with the divisor $D_{i}: z_{i}=0$.

The relation obtained between first Chern classes should hold for any twistor space considered for the compactification where we intend to use two-four split. Thus, in this subsection, even though 2-4 split of tangent space followed by change in almost complex structure gives a vanishing first Chern class for $\mathbb{CP}^3$ case, one cannot perform similar modifications
on any general smooth toric manifold. 


\subsection{More about 1-form and holomorphic 3-form}\label{rigid}

In this section, we study 1-form K which plays a central role in SU(3)-structure we are considering from the 6-manifold perspective\cite{Grana:2005a,LT:2009b}.

Using the Chern classes for modified almost complex structure from previous subsection 4, We know $c_1^{new}=0$. Let's understand more about top-class.
\begin{align}
c(T^{(1,0)}M)=(1-c_1)(1+c_1+c_2)=1+c_{1}^{new}+c_{2}^{new}+c_{3}^{new}
\end{align}
This gives $c_{3}^{new}= -c_1\times c_2$ and $c_{2}^{new}= -c_{1}^{2}+c_2$.
Let's compute the top class of holomorphic cotangent bundle twisted with a line bundle using equations from Appendix A.
\begin{align}
c_3(T^{*(1,0)}\otimes L)&= y^3+c^{new}_{2}(T^{*(1,0)})y+c^{new}_3(T^{*(1,0)})\nonumber\\ 
&=y^3+c^{new}_{2}(T^{(1,0)})y-c^{new}_3(T^{(1,0)})\\
c_3(T^{*(1,0)}\otimes L)&= y^3-(c_1^{2}-c_2)y+c_1\times c_2
\end{align} 
Now we should ask whether it is possible to have a non-vanishing holomorphic section of this twisted bundle.
Firstly, we should see what happens in $\mathbb{CP}^3$ case.  It has $c_{1}(T^{(1,0)})=2g, c_{2}(T^{(1,0)})=2g^2$ for tangent bundle from section 4.1, thus $c_3(T^{*(1,0)}\otimes L)= y^3-(4g^{2}-2g^2)y+4g^3$. In order to have $c_3(T^{*(1,0)}\otimes L)=0$ in $\mathbb{CP}^3$, $y=-2g$, there exists a solution for y. 
In general, it is important to observe that $y=-c_1$ is a solution and such line bundle is easy to find out for various smooth toric manifolds. We observe that the 1-form needed in the construction explained in section 3.1 should be obtained as mentioned above with a proper choice of line bundle. Furthermore, it is interesting to observe that $c_{1}(T^{*(1,0)}M_{2}\otimes L)=c_{1}(T^{*(1,0)}M_{2})+c_{1}(L)=-c_{1}(T^{(1,0)}M_{2})+y=c_{1}-c_{1}=0$. Thus, the non-vanishing holomorphic section of such a twisted bundle is associated with the 1-form, one uses for wedging it non-vanishing 2-form on dual of $\xi$. This fact is always known, but here we say how to obtain such form with $T^{*}_{2}$.

In this section, we showed that with the 2-4 split of tangent space and flipped sign of almost complex structure in 2 of those directions leads to the fact that one can twist the cotangent bundle with appropriate line bundle and one obtains non-vanishing holomorphic 1-form which one can use later for getting nowhere vanishing holomorphic 3-form. This is always understood through supersymmetry conditions, but here we obtain the proper understanding using Chern classes. This picture is known more or less, we give a better procedure to obtain an 1-form which can keep vanishing 1st Chern class following section 3.1.


\section{Local Analysis for SU(3) structure}\label{rigid}

In this section, we study the conditions to construct SU(3) structure on Toric manifolds explained in Section 3 in order to obtain massive Type IIA flux vacua. 


\subsection{$AdS_{4}$ flux vacua in Type IIA theories}
In this section, we will study the 4d flux compactification of (massive) Type IIA supergravity on SU(3) structure manifolds.
Bosonic fields of massive IIA theory are a metric $g_{\mu\nu}$, an RR 1-form potential A and 3-form potential C, a NSNS 2-form potential B and a dilaton $\phi$.
In this note, we are interested in the supersymmetric vacua from 10d point of view.  We will consider a 10 dimensional background, a warped product of four dimensional space and an internal six dimensional manifold.
\begin{equation}
ds^2_{10d} = e^{2A(y)}g_{\mu\nu} dx^{\mu}dx^{\nu} + ds^{2}_{6} 
\end{equation}
In order to preserve the symmetry of 4d space-time, fluxes needs to be chosen appropriately.

In massive Type IIA, $AdS_{4}$ vacuum can be obtained using following choice of internal fluxes, (We will follow conventions of \cite{Tomasiello:2007a} for our discussion):
\begin{eqnarray}
&H=2mRe\Omega \nonumber\\
&g_{s}F_{6}=-\frac{1}{2}\tilde{m}J^3 \nonumber\\
&g_{s}F_{4}=\frac{3}{2}mJ^2 \nonumber\\
&g_{s}F_{2}=-W_{2}^{-}+\frac{1}{3}\tilde{m}J\nonumber\\
&g_{s}F_{0}=5m
\end{eqnarray}
and 
\begin{eqnarray}
&dJ=2\tilde{m}Re\Omega \nonumber\\
&d\Omega=i(-\frac{4}{3}\tilde{m}J^2+W_{2}^{-}\wedge J)
\end{eqnarray}

Also, $3A=\phi=$ constant. The cosmological constant is given by $\Lambda =-3(m^2+\tilde{m}^2)$.
To obtain all equations of motion, supersymmetric equations have to be complemented with Bianchi identities for fluxes. For $F_{n}$, bianchi identity is $dF_{n}=H\wedge F_{n-2}+Q\delta(sources)$ where the source contribution comes from D-branes or O-planes.\\ 

Let's restrict our discussion to the sourceless case, $dF_{n}=H\wedge F_{n-2}$. The only complication occurs when $n=2$, which puts a restriction on $W_{2}^{-}$
\begin{eqnarray}
dW_{2}^{-}=(\frac{1}{3}\tilde{m}^2-5m^{2})2Re\Omega
\end{eqnarray}
To obtain $AdS_{4}$ flux vacuum without localised sources, it is important to satify all equations (16),(17) and (18). One should notice that massive Type IIA solutions have $W_{3}=W_{4}=W_{5}=0$.

Since this setup and corresponding solution of massive Type IIA were achieved for $\mathbb{CP}^3$ with the help of almost complex structure explained in Section 4.1, we should try to see whether we can obtain this setup for other possible smooth Toric manifolds.

\subsection{Analysis}\label{rigid}
Let's perform the local analysis of differential system which K satisfies. We assume that smooth toric variety is chosen such that it satisfies the condition from section 4.2. Choose coordinates such that $z^{i}=e^{t^{i}}$. In new coordinates, vector fields look like $V^{a}=\sum_{i}Q^{a}_{i}\partial_{t_{i}}$ and $K=K_{i}dt^{i}$.

From condition (a),
\begin{equation}
\sum_{i} Q_{i}^{a}K_{i}=0
\end{equation}
Using Cartan's magic formula, condition (b) can be simplified further.
\begin{align}
\mathcal{L}_{ImV^a}K&=(d\circ \imath_{ImV^{a}}+\imath_{ImV^{a}}\circ d)K\nonumber\\
&=\imath_{ImV^{a}}(dK)\nonumber\\
&=\frac{1}{2}\{\sum_{j,j\neq i}Q^{aj}\partial_{j}K_{i}dt^{i}-\sum_{i,j\neq i}Q^{ai}\partial_{j}K_{i}dt^{j} -\sum_{j}Q^{aj}\partial_{\bar{j}}K_{i}dt^{i}\}
\end{align}
One can use condition (a) and it gives locally $\sum_{i}Q^{ai}\partial_{j}K_{i}=0$. Thus we get
\begin{equation}
\mathcal{L}_{ImV^a}K=\frac{1}{2}\{\sum_{j}Q^{aj}(\partial_{j}K_{i} -\partial_{\bar{j}}K_{i})dt^{i}\}
\end{equation}

Let's understand the eigenvalue relation of condition (b) component-wise in these coordinates.
\begin{equation}
\frac{1}{2}\{\sum_{j}Q^{aj}(\partial_{j}K_{i} -\partial_{\bar{j}}K_{i})\}=\frac{1}{4} (\sum_{k}Q^{a}_{k})K_{i}
\end{equation}
Suppose $K_{i}=f\cdot G_{i}$ such that f is given by $\sum_{j}Q^{aj}(\partial_{j}-\partial_{\bar{j}})f=(\frac{1}{2}\sum_{k}Q^{a}_{k})f$. Thus, f will be of form $e^{\lambda \cdot \bar{t}}$ such that $\sum_{j}Q^{aj}\lambda_{j}=\sum_{k}Q^{a}_{k}$. This allows $\lambda$ to take the following form, 
\begin{equation}
\lambda_{j}=-\frac{1}{2}+\frac{1}{2}p(t^{i})\sigma_{j}
\end{equation}
such that $\sigma \in Kernel(Q)$ and $p(t^{i})$ is a complex-valued scalar function.

To have $K_{i}=f\cdot G_{i}$ as the local description, the restriction on $G_{i}$ is following:
\begin{equation}
\sum_{j}Q^{aj}(\partial_{j}-\partial_{\bar{j}})G_{i}=0
\end{equation}
The simplest solution for $K_{i}$ can have is $fG_{i}(Re(t^{j}))$.

\subsection{Changing the Torsion classes}\label{rigid}

In this section, we try to modify SU(3) structures by using K. We would like to see locally if we can find torsion classes for Type IIA flux vacua. 

Natural question to ask is whether we can change the torsion classes for the general smooth, compact toric manifold or not.
Suppose we have K with a $p=0$ in eq. (23) and corresponding real 2-form J and three form $\Omega$ can be computed using eq. (3.5-3.6) with $\alpha = \beta=1 $ and $\gamma=\pi/2$. Let's say this situation leads to 
\begin{eqnarray}
dJ^{old}=\frac{3}{2}Im(\bar{W}^{old}_1\Omega^{old})+W^{old}_{4}\wedge J^{old}+W^{old}_3\\
d\Omega^{old}=W_{1}^{old}{J^{old}}\wedge J^{old}+W^{old}_{2}\wedge J^{old}+\bar{W}^{old}_5\wedge\Omega^{old}
\end{eqnarray}
Now, the goal is to modify K with the help of eq. (23). $K^{new}=e^{p\sigma_{i}Imt^{i}}K^{old}=e^{p\Sigma}K^{old}$.
In this section, we keep $\alpha, \beta^2$ and $\gamma$ as real, gauge invariant functions on the toric variety, but $p(t^{i})$ is purely imaginary function, this choice is made just to have compatible J and $\omega$ as explained in section 2.1 .
\begin{eqnarray}
&J^{new}=\alpha (J_{FS}-\frac{i}{2}K^{old}\wedge K^{*old})-\frac{i\beta^{2}}{2}(K^{old}\wedge K^{*old})\\
&\Omega^{new}=\alpha\beta e^{i\gamma +p^{*}\Sigma}\Omega^{old}
\end{eqnarray}

Let's compute torsion classes in this case.
\begin{align}
&dJ^{new}=\frac{1}{2}[d(\alpha-\beta^2)\wedge J_{FS}+d(\alpha+\beta^{2})\wedge J^{old}+(\alpha+\beta^{2})dJ^{old}]\\
&d\Omega^{new}=d\ln(\alpha\beta e^{i\gamma + p^{*}\Sigma})\wedge \Omega^{new}+ \alpha\beta e^{i\gamma+p^{*}\Sigma}d\Omega^{old}
\end{align}

Using eq. (25) and (26), we get
\begin{align}
dJ^{new}&=\frac{(\alpha + \beta^{2})}{2}\{(\frac{3}{2}Im(\bar{W}^{old}_1\Omega^{old})+W^{old}_{4}\wedge J^{old}+W^{old}_3)\}+\frac{1}{2}d(\alpha-\beta^{2})\wedge J_{FS}\nonumber\\
&+\frac{1}{2}d(\alpha +\beta^{2})\wedge J_{old}\\
&=\frac{(\alpha + \beta^{2})}{2}(\frac{3}{2}Im(\bar{W}^{old}_1\Omega^{old})+(W^{old}_{4}\wedge J^{new}+\frac{d(\alpha+\beta^2)}{\alpha+\beta^2})+\frac{\alpha+\beta^2}{2}W^{old}_3\nonumber\\
&-\frac{(\alpha-\beta^{2})}{2}d\ln(\alpha+\beta^2)\wedge J_{FS} +\frac{1}{2}d(\alpha -\beta^{2})\wedge J_{FS}-\frac{1}{2}(\alpha-\beta^2)W_{4}^{old}\wedge J_{FS}\\
d\Omega^{new}&=d\ln(\alpha\beta e^{i\gamma + p^{*}\Sigma})\wedge \Omega^{new}+\alpha\beta e^{i\gamma+p^{*}\Sigma}d\Omega^{old}\\
d\Omega^{new}&=[d\ln(\alpha\beta e^{i\gamma + p^{*}\Sigma})+\bar{W}^{old}_{5}]\wedge \Omega^{new}+ \frac{\alpha \beta e^{i\gamma+p^{*}\Sigma}}{(\alpha +\beta^2)^2}W_{1}^{old}[J_{new}\wedge J_{new}-2\beta^{2}J_{FS}\wedge J_{new}\nonumber\\
&+\beta^{4}J_{FS}\wedge J_{FS}]+\frac{\alpha \beta e^{i\gamma+p^{*}\Sigma}}{(\alpha +\beta^2)}[W_{2}^{old}\wedge J_{new}-W_{2}^{old}\wedge J_{FS}]
\end{align}

Now, we will see how to change the torsion classes and restrictions associated with the change.
\begin{align}
W_{5}^{new}&=W_{5}^{old}+d\ln(\alpha\beta e^{-i\gamma + p\Sigma})\\
W_{4}^{new}&=W_{4}^{old}+d\ln(\alpha +\beta^{2})\\
W_{3}^{new}&=\frac{1}{2}(\alpha +\beta^{2})W_{3}^{old}+\frac{d(\alpha-\beta^2)}{2}\wedge J_{FS}+\frac{(\alpha-\beta^2)}{2}W_{4}^{old}\wedge J_{FS}\nonumber\\
&-\frac{\alpha -\beta^2}{2}d\ln{(\alpha+\beta^2)}\wedge J_{FS}
\end{align}

Since we are working with toric manifolds, we know that $H_{1}(M)=0$, thus we know that if one form is closed, then it is exact. Suppose $W_{4}$ is exact, so we can choose function $(\alpha+\beta^2)$ such that eq.(36) gives $W_{4}^{new}=0$. With this condition, eq.(37) gives $$W_{3}^{new}=\frac{1}{2}(\alpha +\beta^{2})W_{3}^{old}+\frac{d(\alpha-\beta^2)}{2}\wedge J_{FS}$$ 
$W_3$ is a primitive (1,2)+(2,1) form, in order to enforce primitivity, one option is to impose $\alpha=\beta^2$. We have a function $\alpha$  to adjust $W_4$ to zero locally, one does not have enough functions with the chosen ansatz to tune $W_{3}$ to zero. This ansatz might be more useful in finding classical dS solutions mentioned in \cite{vRiet:2010}. \\
The general idea on the lines of Calabi-Yau compactifications (Calabi conjecture) is to understand global properties with topological conditions and find a solution locally. Here we see that for arbitrary toric case, we cannot find massive Type IIA solution locally.
Similarly, we can change the $W_5$ by adding an exact form with the help of $\sigma$ of eq. (23). This process does change $W_{1}$ and $W_{2}$ beyond multiplying by functions, but one has to fix coefficients appropriately. In general, we have $\alpha,\gamma,\sigma,p$ and freedom in $J_{FS}$ to adjust $W_{i}$. We have shown that one can tune torsion classes on case-by-case basis and in general when $W_4$ is closed. At this stage, we can hope to find more solutions of massive Type IIA by adjusting $W_{i}$ suitably for smooth toric manifolds when conditions above are matched.

\section{Discussions}\label{Discussions}
We showed that  first Chern class computations for new almost complex structure can vanish on smooth toric manifolds. We also had to study top Chern class properties for using this construction which played an important role in getting nowhere vanishing holomorphic 3-form. 
After understanding global topological conditions, we carried out local analysis of differential system and showed that it is possible to change torsion classes associated with the SU(3) structures. Here we are trying to find more Type IIA flux vacua.  We would like to conclude that in certain cases, torsion classes can be changed appropriately, but there is no explicit argument for the class of toric manifolds in general. Type IIA vacua obtained in such cases would be $AdS_{4}$. One might find this technique useful to explore classical dS solutions\cite{vRiet:2010} with smooth, compact toric manifolds\footnote[2]{Work In Progress.}. 

\section*{Acknowledgments}
I would especially like to thank Michael R Douglas for regular guidance during the project and John W Morgan for insightful mathematical discussions. I would also like to thank Anant Atyam, Peng Gao, Daniel Park, Anibal Medina for discussions. I would like to thank Thomas Van Riet, Magdalena Larfors and Andrew Frey for pointing out useful works in this context. This research was supported in part by DOE grant DE-FG02-92ER40697.\\

\appendix

\section{Chern Classes}\label{details}

Here we will discuss Chern classes of vector bundles and their properties in brief\cite{Nakahara:1990}.\\
{\bf{Definition:}} Let $E \rightarrow M$ be a complex vector bundle whose fiber is $\mathbb{C}^k$. The structure group G ($\subset GL(k,\mathbb{C})$) with a connection $\mathcal{A}$ and its strength $\mathcal{F}$, then the Chern class is defined as
\begin{equation}
c(E)=det(1+\frac{i\mathcal{F}}{2\pi})
\end{equation}
It can be decomposed as
\begin{equation}
c(E)=1+c_{1}(E)+c_{2}(E)+...
\end{equation}
such that i-th Chern class $c_{i}(E)\in \Omega^{2i}(M)$. Hence, it is clear that if n is the rank of a vector bundle E, then $c_{i}(E)=0$ for $i >n$.  \\
Properties of Chern Class: For vector bundles E and F, tangent bundle T and cotangent bundle $T^{*}$ and line bundle L,\\
\begin{eqnarray}
c(E\oplus F)=c(E)\wedge c(F)\\
c(L)=(1+x)\\
c(E\otimes L)=\sum_{i=1}^{n}c_{i}(E)(1+x)^{n-i}\\
c(T^{*})=\sum_{k}(-1)^{k}c_{k}(T)
\end{eqnarray}

\section{Toric Geometry}\label{details}

We have considered the symplectic description of toric varieties for this work. Toric geometry can be described using simple combinatorial data. Interested reader can follow \cite{Fulton:1993ab, Cox:2009}.

Consider a rank-d integer lattice $N \cong Z^d$ and the real extension of N, $N_\mathbb{R}= \mathbb{R} \otimes N$. A subset $\sigma \subset N_\mathbb{R}$ is a called a strongly convex rational polyhedral cone with apex 0 if $\sigma \cap (-\sigma)={0}$ and there exist elements $v_1$,...,$v_r$ of N such that
\begin{equation} 
\sigma=\{a_{1}v_1+...a_{r}v_r; 0\leq a_1,...a_r \in \mathbb{R}\}.
\end{equation}
The set $v_1$,...,$v_r$ is usually called generators of cone $\sigma$. $\tau$ is called a face of $\sigma$ if its generators are a subset of the generators of $\sigma$. 

A fan $\Sigma$ is a collection of cones $\{\sigma_1,...,\sigma_k\}$ such that 
\begin{description}
 \item[1.] $\sigma \in \Sigma$ is a strongly convex rational polyhedral cone.
 \item[2.] If $\sigma \in \Sigma$ and $\tau$ is a face of $\sigma$, then $\tau \in \Sigma$.
\item[3.] If $\sigma, \sigma' \in \Sigma$, then $\sigma \cap \sigma'$ is a face of both cones.
\end{description}
The support $|\Sigma|$ of a fan $\Sigma$ is the union of all cones in the fan.

The toric variety X($\Sigma$) can be constructed corresponding to a fan $\Sigma$ by taking the union of affine toric varieties. \cite{Fulton:1993ab} One can obtain lot of information about toric manifolds using the fan description.

A Weil divisor is a finite sum of irreducible hypersurfaces with a co-dimension one. $D = \sum n_i V_i$ such that $n_{i}\in \mathbb{Z}$ and $V_{i}$ are irreducible subvarieties. There is a one-to-one mapping from generators of $\Sigma(1)$ and T-Weil divisors. If $\{v_1,...,v_k\}$ are rays in a fan, Weil divisor is $$D=\sum_{i=1}^{k}a_{i}V_{i}$$ where $a_{i}$ are integers. For toric varieties, there is a correspondence between divisors and line bundles.

One of the interesting property of smooth toric manifolds for flux compactifications is all odd betti numbers vanish and even betti numbers are given by
\begin{equation}
\beta_{2k}=\sum_{i=k}^{n}(-1)^{(i-k)}\binom{i}{k} d_{n-i}
\end{equation}
Here $d_{k}$ is the number of k-dimensional cones in $\Sigma$.

\end{document}